\documentclass[seceq]{ptptex}

\usepackage{graphicx}
\usepackage{amsmath,amssymb,amsxtra}
\usepackage{float}

\preprintnumber[3cm]{
OCU-PHYS-248\\AP-GR-34\\PTPTeX ver.0.8\\ \today}

\title{
Can Inhomogeneities Accelerate the Cosmic Volume Expansion?
}

\author{
$^1$Tomohiro \textsc{Kai}, $^1$Hiroshi \textsc{Kozaki}, 
$^1$Ken-ichi \textsc{Nakao}, 
$^2$Yasusada \textsc{Nambu} \\
and $^1$Chul-Moon \textsc{Yoo}
}

\inst{

$^1$Department of Physics,~Graduate School of Science,
~Osaka City University \\ Osaka 558-8585,~Japan\\
and\\
$^{2}$Department of Physics, Graduate School of Science, 
Nagoya University, \\
 Nagoya 464-8602, Japan \\
(Recieved May 22, 2006)
}

\date{\today}

\abst{
If expanding and contracting regions coexist in the
universe, the speed of the cosmic volume expansion can be accelerated. 
We construct simple inhomogeneous dust-filled universe models in which 
the speed of the cosmic volume expansion is accelerated for finite periods. 
These models are constructed by removing spherical domains from the 
Einstein-de Sitter universe and filling each domain with a 
Lema\^{\i}tre-Tolman-Bondi dust sphere possessing the same gravitational 
mass as the removed region. 
This represents an exact solution of the Einstein equations. 
We find that acceleration of the 
cosmic volume expansion is realized in some cases when the size of the 
contracting region is comparable to the 
horizon radius of the Einstein-de Sitter universe though this model is very different from the universe 
observed today. This result implies 
that non-linear general relativistic effects 
of inhomogeneities are very important to realize the acceleration of the cosmic volume expansion. 
}
\begin{document}

\maketitle


\section{Introduction}


The accelerating expansion of the universe indicated by 
observational data for the luminosity distance of Type Ia supernovae
and the cosmic microwave background radiation is a great mystery in 
modern cosmology. The acceleration of a homogeneous and isotropic
Friedmann-Robertson-Walker (FRW) universe implies the existence
of a dark energy component of matter fields which violates the energy
conditions and whose nature is unknown.  Conversely,
if the universe is not homogeneous and isotropic, 
the observational data do not necessarily indicate an accelerating 
expansion of the universe, or even if the cosmic expansion is 
accelerating, it 
does not necessarily imply the existence of dark energy. Thus, 
to account for the observational data  
without introducing dark energy, arguments concerning 
the effects of inhomogeneities have been made. 
Roughly speaking, there are two such arguments. 
One is that the apparent acceleration 
of our universe can be regarded as a result of  
an almost spherically symmetric but inhomogeneous peculiar velocity field,  
assuming that we are located in the vicinity of its symmetry 
center.\cite{CelerierM:AA353:2000,TomitaK:MNRAS326:2001,IguchiH:PTP108:2002,VanderveldRA:0602476:2006} 
With this argument, the acceleration of the cosmic volume 
expansion is not necessary. The other 
argument is that the apparent acceleration of our universe results 
from backreaction effects due to inhomogeneities in the background 
FRW universe.\cite{KolbEW:0503117:2005,KolbEW:0506534:2005}
In spite of its viability, the former argument is simple and clear. 
By contrast, at present it is unclear whether the backreaction 
effects of inhomogeneities can actually accelerate the 
cosmic volume expansion and, further, account for the observed distance-redshift 
relation.\cite{FlanaganEE:PRD71:2005,HirataCM:PRD72:2005,IshibashiA:CQG23:2006}. 

Perturbative analysis of 
backreaction effects on the background FRW universe shows that spatial
inhomogeneities behave as a positive spatial curvature, and their presence reduces
the expansion rate of the FRW
universe\cite{RussH:1997,NambuY:PRD62:2000,KozakiH:PRD66:2002.1}.
Further, recently Kasai et al. conjectured that a
non-linear backreaction cannot accelerate the cosmic volume expansion
even if the inhomogeneities are very large\cite{KasaiM:0602506:2006}.
By contrast, one of the present  authors,
Nambu, and a collaborator, Tanimoto, pointed out the possibility that 
the non-perturbative features of inhomogeneities are necessary to 
realize an accelerating expansion of the universe 
through backreaction effects.\cite{NambuY:0507057} 
They proposed a model of the universe 
containing both expanding and contracting regions and showed that a spatially
averaged scale factor defined in terms of the volume of a spatial
domain can exhibit accelerated expansion if the size of each region is 
properly chosen. 

At present, it is still unclear whether the acceleration of the 
cosmic volume expansion could be realized through 
the effects of inhomogeneities in the universe without introducing 
dark energy. Thus this point must be investigated before we attempt 
to explain the observed distance-redshift relation with this approach. 
As Nambu and Tanimoto have not presented a correct example of 
an accelerating universe, it is necessary 
to determine whether their mechanism actually works. 
In this paper, in order to do this, we study 
the comoving volume of a highly inhomogeneous dust-filled universe. 

This paper is organized as follows. In \S 2, we give a 
qualitative description of a
scenario in which inhomogeneities lead to the  acceleration of the
cosmic volume expansion. In \S 3, in order to obtain quantitative
information about the acceleration of the cosmic  volume expansion, we
study the stuffed Swiss-cheese model. Section 4 is devoted to
a summary and discussion. In this paper, we employ units in which $c=G=1$ and
basically follow the conventions for the Riemann and metric tensors and
the notation used in the textbook of Hawking and
Ellis\cite{HawkingSW:CUP:1973}.

\section{One scenario, a qualitative argument}

\begin{figure}[H]
  \centering
  \includegraphics[width=0.6\linewidth,clip]{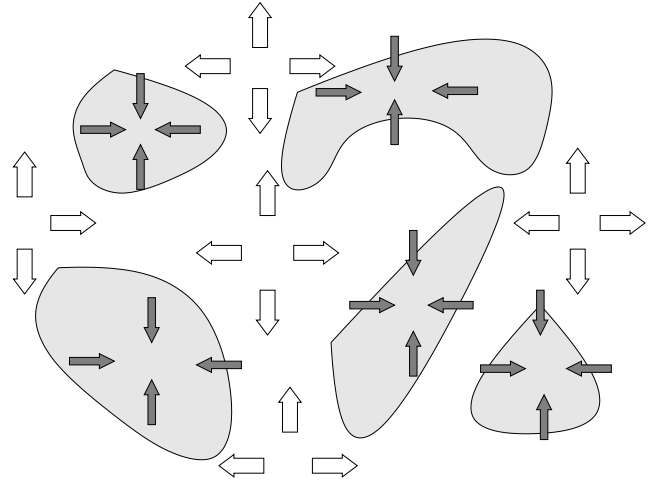}
  \caption{An inhomogeneous universe with expanding and contracting
    regions.}
  \label{fig:inhomo}
\end{figure}

We consider an inhomogeneous dust-filled universe which  is initially
expanding everywhere.  We assume that at some stage, 
the dust begins contracting in
some domains and continues expanding elsewhere 
(see Fig.~\ref{fig:inhomo}).
Such a situation is consistent with our conventional picture of the real
universe: Structures (stars, galaxies, etc.)   
form within contracting domains, 
or there might be large-scale bulk velocity fields due to large mass concentrations, 
in an overall expanding universe.
Further, we assume that the universe is almost periodic and thus
homogeneous in an average sense. We adopt a dust-comoving Gaussian
normal coordinate system in which the line element is given by
\begin{equation}
ds^2=-dt^2+\gamma_{ij}(t,x^k)dx^i dx^j ,
\end{equation}
where $i,j,k=1,2,3$ represent the spatial components, and 
$\gamma_{ij}(t,x^k)$ is the induced metric on the 
spacelike hypersurface $t=$[constant],
which is orthogonal to the trajectories of dust particles. 
The stress-energy tensor of dust is given by
\begin{equation}
T^{\mu\nu}=\rho u^\mu u^\nu,
\end{equation}
where $\rho$ is the rest mass density of dust and $u^\mu=\delta_0^\mu$ 
is the 4-velocity of dust particles. 

We consider a compact spatial domain $D$ in each $t=$[constant] hypersurface 
by assuming that $D$ is dust-comoving. By definition, $D$ 
contains a fixed rest mass of dust.  We assume
that the domain $D$ is so large that the spatial periodicity  is 
recognizable in this domain. The volume $V_D$ of $D$  is defined by
\begin{equation}
V_D=\int_D\sqrt{\gamma}\,d^3x,
\end{equation}
where $\gamma$ is the determinant of the spatial metric $\gamma_{ij}$.
Following Refs. 13) and 16), we define
the spatially averaged scale factor $a_D(t)$ of the domain by
$3{\dot{a}_D}/{a_D}={\dot{V}_D}/{V_D}$,  where the dot represents
the derivative with respect to $t$.  This definition is equivalent to the relation
\begin{equation}
a_D(t) \propto V_D^{1/3}(t).
\end{equation}

Because we assume that the expanding and contracting domains coexist within 
the domain $D$, we rewrite the volume $V_D$ in the form 
\begin{equation}
V_D=V_{\rm e}+V_{\rm c},
\end{equation}
where $V_{\rm e}$ and $V_{\rm c}$ are the comoving volumes  
of the union of the expanding domains and that of the contracting 
domains, respectively. The time derivative of the volume is given by 
\begin{equation}
\dot{V}_D
=\dot{V}_{\rm e}+\dot{V}_{\rm c}
=|\dot{V}_{\rm e}|-|\dot{V}_{\rm c}|.
\end{equation}
We see from the above equation that even though $\dot{V}_D$ is
initially  positive, once $\dot{V}_{\rm c}$ becomes dominant,
$\dot{V}_D$ can be small or even negative. In other words, due to
the appearance of  contracting domains in $D$, the cosmic volume
expansion of $D$  can greatly slow down, or the volume $V_D$ of $D$
can decrease.  However, the contracting domains will collapse, and then
their contributions to the volume $V_D$ will necessarily become
negligible; in realistic situations, almost stationary  structures,
like stars, black holes and   galaxies, are formed. This means
that $\dot{V}_{\rm e}$ will be  dominant in $\dot{V}_D$, and thus the volume
$V_D$ begins to increase again. Here it  should be noted that 
acceleration of the volume expansion is  realized near the end of the
slowdown or contraction stage. During this period,  the second-order time
derivative of the effective scale factor $a_D(t)$  becomes positive. 
This is the mechanism of the acceleration of the cosmic volume  expansion
proposed by Nambu and Tanimoto\cite{NambuY:0507057}.

\section{Stuffed Swiss-Cheese Model}

In order to quantitatively study the effects of inhomogeneities in the
universe, we construct a specific model.  
First, we consider the Einstein-de Sitter universe (EdS)  and remove
disconnected spherical domains from it, so as to guarantee spatial
periodicity. Next, each removed region is  stuffed with 
spherically symmetric but inhomogeneous dust whose  Misner-Sharp
mass\cite{MisnerCW:PR136:1964} is the same as that of the original EdS regions.  
The dynamics of the inhomogeneous dust ball are described by the so-called
Lema\^{\i}tre-Tolman-Bondi (LTB) solution.  We call this model the
stuffed Swiss-cheese (SSC) model.  We assume that the LTB regions
are initially expanding   but eventually begin to contract and form
singularities  and black holes.

The line element of the LTB region is given by 
\begin{equation}
 ds^{2}=-dt^{2}+\frac{Y'{}^{2}(t,\chi)}{1-k(\chi)\chi^{2}}d\chi^{2}
        +Y^{2}(t,\chi)(d\theta^{2}+\sin^{2}\theta d\varphi^{2}), 
\label{eq:line-element}
\end{equation}
where the prime  denotes  differentiation  with respect to
the radial coordinate \(\chi\),  which is assumed to be non-negative.  This
coordinate system is a dust-comoving Gaussian normal coordinate system, as  
in the previous section. The  Einstein equations lead to the equations for 
the areal radius \(Y(t,\chi)\)  and the rest mass density
\(\rho(t,\chi)\) of the dust,
\begin{eqnarray}
 \dot{Y}^{2} &=& -k(\chi)\chi^{2} + \frac{2M(\chi)}{Y},\label{eq:einstein}  \\
 \rho &=& \frac{M'(\chi)}{4\pi Y'Y^{2}},\label{eq:density}
\end{eqnarray}
where \(k(\chi)\) and \(M(\chi)\) are arbitrary functions of the
radial coordinate.  We assume that $\rho$ is non-negative and
$Y$ is monotonic with  respect to $\chi$, i.e., $Y'>0$ in the
regular region.   This assumption leads to $M'\geq0$, and thus we can set
\begin{equation}
M(\chi)=\frac{4\pi\rho_{0}}{3}\chi^{3}, \label{eq:mass-form}
\end{equation}
where $\rho_{0}$ is a non-negative arbitrary constant.  Our treatment does not 
lose generality with the above choice of \(M(\chi)\).  Equations
(\ref{eq:line-element})--(\ref{eq:density}) are invariant under the
rescaling of the radial coordinate \(\chi\),
\begin{equation}
 \chi \rightarrow \tilde{\chi}=\tilde{\chi}(\chi).
\end{equation}
By virture of this property, the above form of $M(\chi)$ is entirely general. 

The solution of Eq.~\eqref{eq:einstein} is given by
\begin{eqnarray}
 Y&=&\frac{4\pi\rho_{0}}{3k(\chi)}\left(1-\cos\left(\sqrt{k(\chi)}\,\eta\right)\right)\chi
, \label{eq:k>0}\\
 t-t_{\rm  i}(\chi)&=&\frac{4\pi\rho_{0}}{3k(\chi)}\left(\eta
   -\frac{1}{\sqrt{k(\chi)}}\sin\left(\sqrt{k(\chi)}\,\eta\right)\right),
\label{eq:t-solution}
\end{eqnarray}
where \(t_{\rm i}(\chi)\) is an arbitrary function.  We can use this
form of the solution for any sign of $k(\chi)$. Note that
\(t_{\rm i}(\chi)\) is the time at which a shell focusing singularity
appears, where `shell focusing singularity' means $Y=0$ for $\chi>0$
and $Y'=0$ at $\chi=0$.  In this paper, we consider the region satisfying
$t>t_{\rm i}$, and hence the time $t=t_{\rm i}$ corresponds to the Big
Bang singularity. Hereafter, we assume a simultaneous Big Bang, 
i.e., $t_{\rm i}=0$.  Using the terminology of cosmological perturbation theory, 
there are only growing modes near the Big Bang singularity in this 
scenario.

We denote the boundary between the LTB region and the EdS region  by
$\chi=\chi_{\rm b}$. Then, we divide the LTB region  into four regions,
$[0,\chi_1)$, $[\chi_1,\chi_2)$,  $[\chi_2,\chi_3)$ and
$[\chi_3,\chi_{\rm b})$.  We use the following spatial profile of the
curvature function:
\begin{equation}
 k(\chi)=
  \begin{cases}
   k_{0} & \text{for~~~} 0\leq\chi<\chi_{1}, \\ 
   \dfrac{k_{0}}{2\chi^{2}}
    \left\{
     \dfrac{(\chi^{2}-\chi_{2}^{2})^{2}}{\chi_{1}^{2}-\chi_{2}^{2}}
     +\chi_{1}^{2}+\chi_{2}^{2}
   \right\} & \text{for~~~} \chi_{1}\leq\chi<\chi_{2}, \rule{0pt}{20pt} \\
   \dfrac{k_{0}}{2\chi^{2}}
   \left(\chi_{1}^{2}+\chi_{2}^{2}\right)
   & \text{for~~~} \chi_{2}\leq\chi<\chi_{3}, \rule{0pt}{26pt} \\
   \dfrac{k_{0}}{2\chi^{2}}\left(\chi_{1}^{2}+\chi_{2}^{2}\right) 
   \left\{
     \left(\dfrac{\chi^{2}-\chi_{3}^{2}}
      {\chi_{\rm b}^{2}-\chi_{3}^{2}}
     \right)^{2}-1
   \right\}^{2}
   & \text{for~~~} \chi_{3}\leq\chi<\chi_{\rm b}, \rule{0pt}{26pt}
  \end{cases}
  \label{eq:region-2-3}
\end{equation}
where $k_{0}$ is constant. 
In order to 
guarantee the relation $1-k\chi^{2}>0$, the following inequality should hold:
\begin{equation}
\kappa:={k_{0}\over2}(\chi_{1}^{2}+\chi_{2}^{2})<1.
\label{eq:kappa-def}
\end{equation}
Since we are interested in the case that the LTB regions contract, 
we assume $k_{0}>0$.

The parameter $\kappa$ is closely related to the value of the comoving volume 
in the LTB region $V_{\rm LTB}$, which is bounded below as
\begin{eqnarray}
V_{\rm LTB}&:=&4\pi\int_0^{\chi_{\rm b}}\frac{Y^2Y'}{\sqrt{1-k\chi^2}}d\chi
=4\pi\left(
\int_0^{\chi_{\rm 2}}
+\int_{\chi_2}^{\chi_{\rm 3}}
+\int_{\chi_3}^{\chi_{\rm b}}
\right)
\frac{Y^2Y'}{\sqrt{1-k\chi^2}}d\chi 
\nonumber \\
&=&
\frac{4\pi}{3\sqrt{1-\kappa}}(Y_3^3-Y_2^3)+4\pi\left(
\int_0^{\chi_{\rm 2}}
+\int_{\chi_3}^{\chi_{\rm b}}
\right)
\frac{Y^2Y'}{\sqrt{1-k\chi^2}}d\chi 
\nonumber \\
&>&
\frac{4\pi}{3\sqrt{1-\kappa}}(Y_3^3-Y_2^3)+4\pi\left(
\int_0^{\chi_{\rm 2}}
+\int_{\chi_3}^{\chi_{\rm b}}
\right){Y^2Y'}d\chi
\nonumber \\
&=&\frac{4\pi(1-\sqrt{1-\kappa})}{3\sqrt{1-\kappa}}(Y_3^3-Y_2^3)
+\frac{4}{3}\pi Y_{\rm b}^3, \label{eq:LTB-volume}
\end{eqnarray}
where, for notational simplicity, we denote $Y|_{\chi=\chi_i}$ by $Y_i$. 
Note that the last term in the above equation, $4\pi Y_{\rm b}^3/3$, is equal to the 
volume of the original EdS region, since the areal radius $Y$ is continuous 
across the boundary between the LTB and EdS regions. Also, 
the first term in the last inequaltiy gives a lower bound 
on the volume increase $\delta V$ due to the inhomogeneous geometry: 
\begin{equation}
\delta V> \frac{4\pi(1-\sqrt{1-\kappa})}{3\sqrt{1-\kappa}}(Y_3^3-Y_2^3).
\end{equation}
In the limit $\kappa\rightarrow1$, the volume increase becomes infinite. 
Thus if $\kappa$ is almost equal to unity, the volume of the LTB region 
can be dominant in the SSC universe, even if $Y_3$ becomes small. 
Further, note that this result is not related to how the LTB region is 
divided into the four regions $[0,\chi_1)$, $[\chi_1,\chi_2)$, 
$[\chi_2,\chi_3)$ and $[\chi_3,\chi_{\rm b})$, if $\kappa$ is sufficiently close 
to unity.  

The physical meaning of the volume increase is the decrease of the 
binding energy in the LTB region. The binding energy $E_{\rm bind}(\chi)$ 
within the spherical domain of the comoving radius $\chi$ is defined 
by the difference between the gravitational mass $M(\chi)$ and the rest mass of 
the dust, i.e., we have 
\begin{equation}
E_{\rm bind}(\chi):=M(\chi)-4\pi\int_0^\chi\frac{\rho Y^2Y'}{\sqrt{1-k\chi^2}}d\chi,
\end{equation}
where the second term on the right-hand side corresponds to the rest mass of the dust 
included within the spherical domain of the comoving radius $\chi$. In the case of 
vanishing $k_0$, the LTB region is identical to 
the original EdS region, and $E_{\rm bind}(\chi)$ vanishes. 
By using Eq. (\ref{eq:density}) and the same manipulation as in Eq. (\ref{eq:LTB-volume}), 
we obtain an upper bound on the binding energy 
$E_{\rm bind}(\chi_{\rm b})$ of the LTB region as
\begin{equation}
E_{\rm bind}(\chi_{\rm b})
<-\frac{1-\sqrt{1-\kappa}}{\sqrt{1-\kappa}}\left[M(\chi_3)-M(\chi_2)\right].
\end{equation}
The limit $\kappa\rightarrow1$ leads to a negatively infinite binding energy. 
In other words, if $\kappa$ is positive, the rest mass of dust in the LTB region 
is larger than that included in the original EdS region. 
An arbitrarily large rest mass can be put into the removed domain 
if $\kappa$ is set sufficiently close to unity. In this sense, the LTB region 
with $\kappa\simeq1$ is a very high density region compared with the original EdS 
region. 


The singularity formed at the origin, $\chi=0$, can be a null naked
singularity, whereas that formed in the domain $\chi>0$ is
spacelike\cite{ChristodoulouD:CMP93:1984,NewmanRPAC:CQG3:1986,JoshiPS:PRD47:1993}.
In the case of the present model, since the innermost region
$[0,\chi_1)$ is a Friedmann universe with a positive spatial
curvature, the singularity formed at $\chi=0$ is spacelike and thus
not naked. We see from Eq.~(\ref{eq:t-solution}) that the singularity
formation time is $t=8\pi^2\rho_0/(3k^{3/2})$. Because $k$ is
monotonically decreasing with respect to $\chi$, the
singularity formation time is monotonically increasing with respect to
$\chi$, and thus no shell crossing singularity is realized in the
present model.  Therefore, singularities formed through
gravitational collapse in the LTB regions are necessarily spacelike in
the present SSC model.  There is no causal influence of spacelike
singularities on the regular domains. This property allows us to
ignore the contribution of singularities in the calculation of the
volume $V_D$ of the domain $D$, and thus we can follow the evolution of
the cosmic volume expansion even after singularity formation.
Further, it should be noted that the present choice of $k(\chi)$
guarantees that the LTB region smoothly connects to the EdS
universe. The line element of the EdS region is given by the form
\begin{equation}
ds^2=-dt^2+a_{\rm e}^2(t)\left[d\chi^2
+\chi^2\left(d\theta^2+\sin^2\theta d\varphi^2\right)\right],
\end{equation}
where 
\begin{equation}
a_{\rm e}(t)=\left(6\pi\rho_0 t^2\right)^{1/3}.
\end{equation}

We assume that the universe has the periodicity of the  comoving
interval, $2l$, in the original Einstein-de Sitter  universe. With this
assumption, a cubic region of comoving edge length $2l$
corresponds to the compact domain $D$ considered in the previous
section.  We define the scale factor $a(t)$ of this  inhomogeneous
universe as
\begin{equation}
a_D(t)\equiv \frac{V_D^{1/3}}{2l}.
\end{equation}

\subsection{One-scale model}
\begin{figure}[H]
  \centering
  \includegraphics[width=0.45\linewidth,clip]{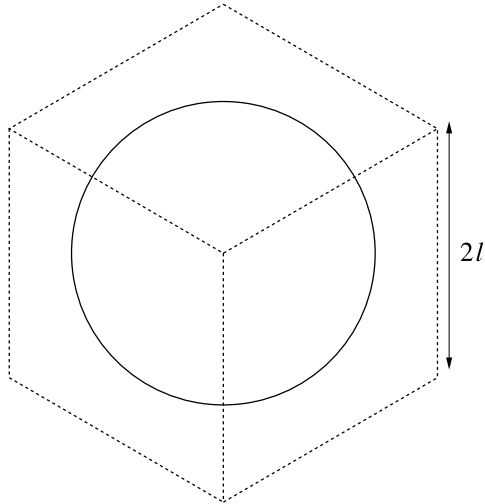}
  \caption{The compact domain $D$ of the one-scale SSC model.}
  \label{fig:one-scale}
\end{figure}

We first consider a model with one LTB region with $\chi_{\rm b}=l$ in
the domain $D$ (see Fig.~\ref{fig:one-scale}).
\noindent
The volume $V_D$ of the domain $D$ is given by
\begin{equation}
V_D=\left(8-{{4\pi}\over3}\right)a_{\rm e}^3(t)l^3
+4\pi\int_0^{a_{\rm e}(t)l}\frac{Y^2dY}{\sqrt{1-k(\chi)\chi^{2}}}. 
\end{equation}
The integral on the right-hand side of the above equation  corresponds
to the comoving volume of the LTB region.  It should be noted that
$\chi$ in this integral  is a function of the areal radius $Y$ at
time $t$ through  Eqs.~(\ref{eq:k>0}) and (\ref{eq:t-solution}). Thus,
this integration  covers the only region which has not yet collapsed
to the singularity,  i.e., the domain satisfying $Y>0$ at time $t$.

\begin{figure}[t]
  \centering
  \includegraphics[width=0.4\linewidth,clip]{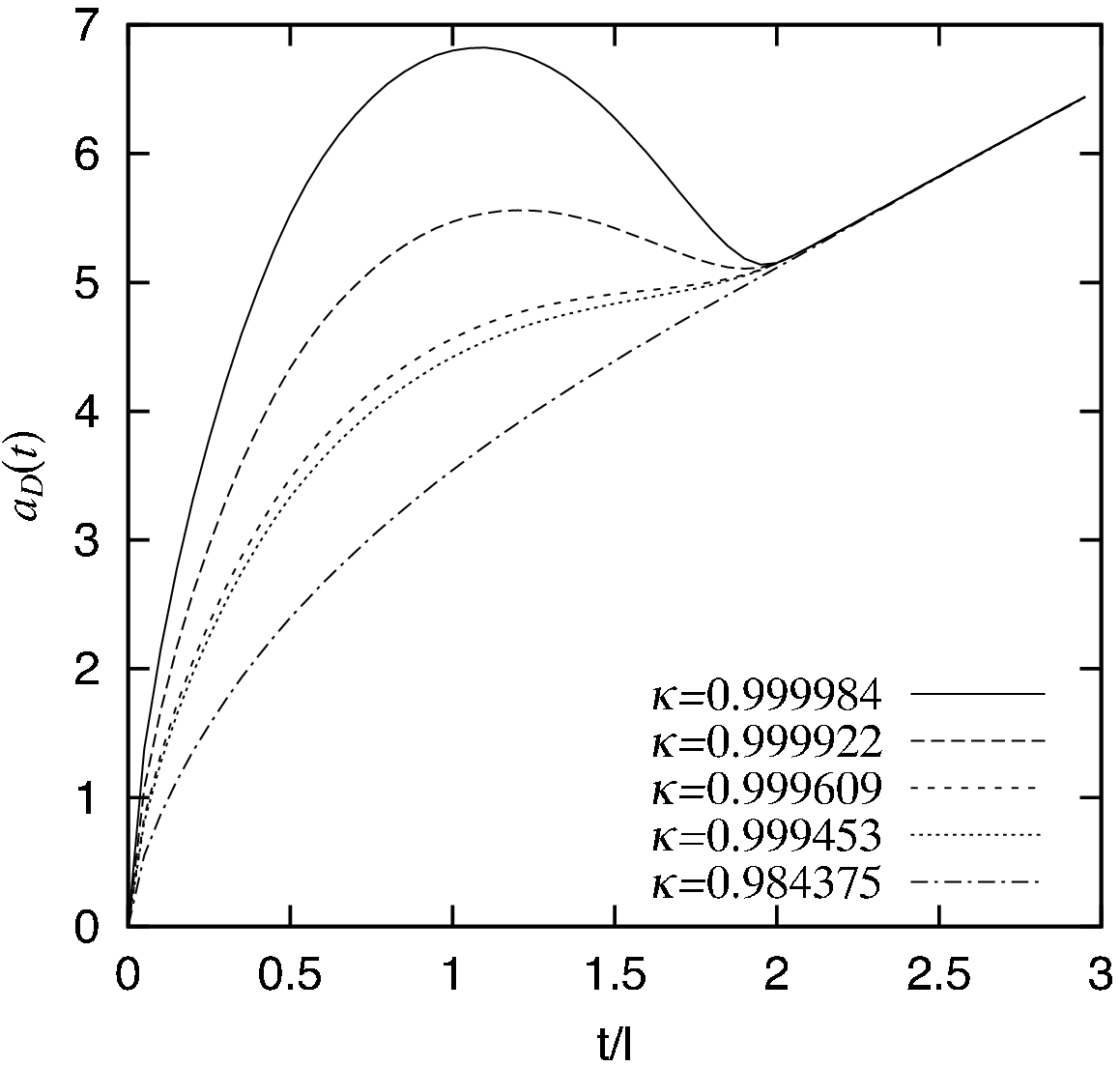}%
  \hspace{0.5cm}
  \includegraphics[width=0.4\linewidth,clip]{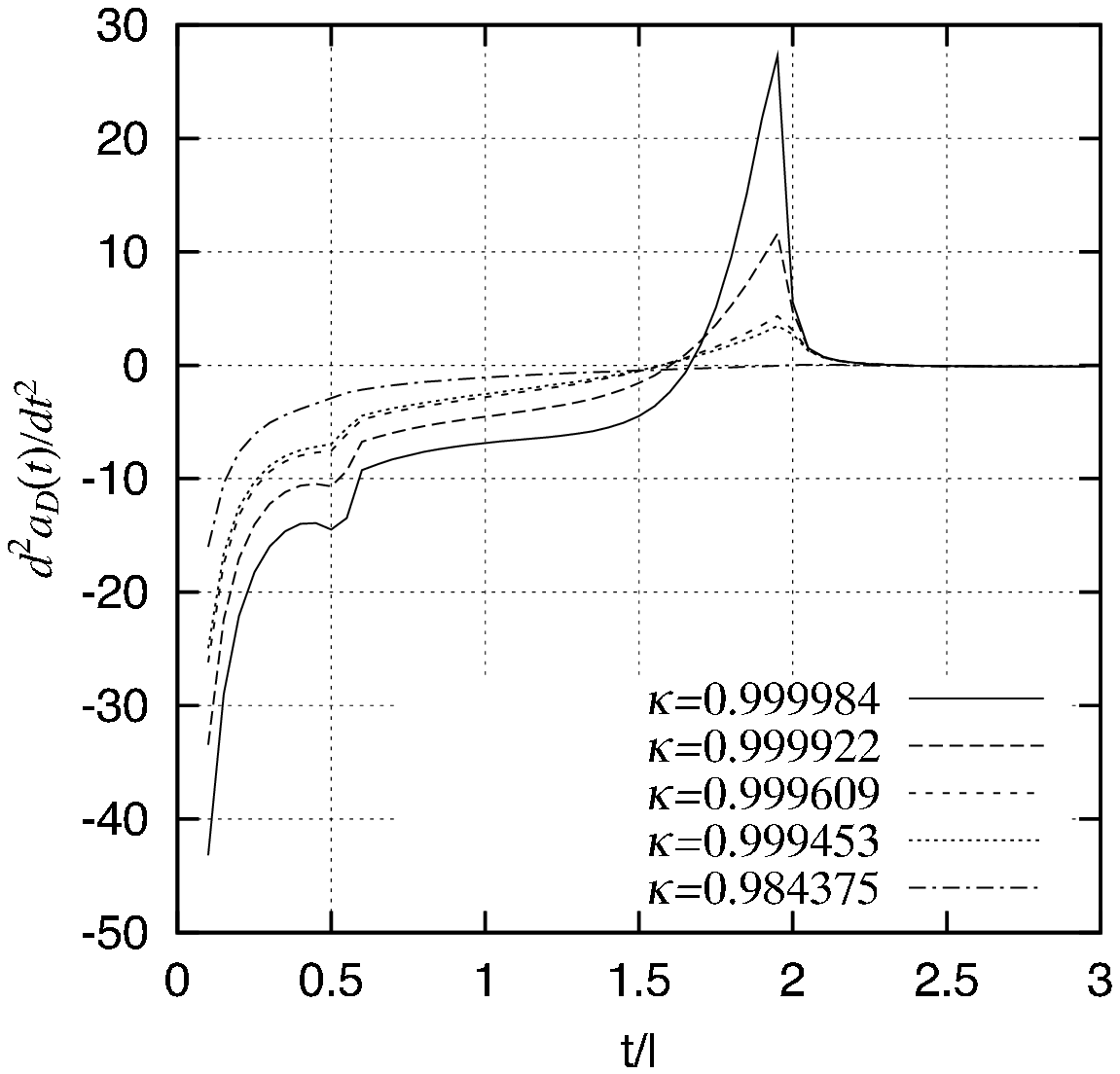}
  \caption{The left panel plots the evolution of the averaged scale factor
    $a_D$ of one scale SSC model for various values of
    $\kappa$.  The right panel plots the acceleration of the averaged scale
    factor $a_D$.}
  \label{fig:expansion-one}
\end{figure}

In Fig.~\ref{fig:expansion-one}, we plot several examples of the
evolution of the scale factor $a_D(t)$. We set $H_0 l=\sqrt{6}/2,
H_0\chi_1=\sqrt{6}/8, H_0\chi_2=\sqrt{6}/4$ and
$H_0\chi_3=3\sqrt{6}/8$ with $H_0=\sqrt{8\pi\rho_0/3}$.  We chose
these values so that the areal radius of the boundary of the LTB
region is equal to the horizon radius of the EdS universe at $t=l$.
These curves represent the scale factors as functions of $t$ for various
values of $\kappa$ defined by Eq.~(\ref{eq:kappa-def}).  As mentioned
in the preceding section, the slowdown of the volume expansion or the
contraction of the domain $D$ actually does occur, due to the contraction of
the LTB region. After the LTB region collapses to a singularity
covered by a horizon and forms a black hole, the volume expansion of
the domain $D$, which has been slowed by the contracting LTB region,
accelerates until the speed of the volume expansion becomes the
same as that of the EdS universe.  Thus, at a time near the end of
the slowdown or contraction stage, an acceleration period appears.
In the cases depicted in Fig.~\ref{fig:expansion-one}, 
acceleration of the volume expansion 
occurs in four cases ($\kappa\geq 0.999453$), and in these four 
cases, volume contraction before the acceleration period 
occurs in the two cases with $\kappa \geq 0.999922$, whereas the other two cases  
with smaller $\kappa$ do not have a contraction period, although  
acceleration does occur. In the case of the smallest 
$\kappa~(=0.984375)$ depicted in Fig.~\ref{fig:expansion-one}, 
no acceleration period appears. 

The necessary condition of the appearance of an acceleration period 
is that the volume of the LTB region becomes dominant in $V_D$. 
As mentioned above, if the parameter $\kappa$ 
is almost equal to unity, the contribution of $\chi$ in the range
$\chi_2\leq\chi<\chi_3$ is dominant in the volume of the LTB region
and, further, makes the volume of the LTB region dominant in $V_D$ 
for a long time before the domain $\chi\leq\chi_3$ completely 
collapses. Thus, it is conjectured that acceleration of the cosmic volume 
expansion is realized if and only if $\kappa$ is almost equal to unity, 
and we can see from Fig.~\ref{fig:expansion-one} that this is in fact 
the case in the present model. 

The size of the LTB region relative to the cosmological horizon is
very important. If there are many contracting LTB regions within the
cosmological horizon of the EdS universe in the acceleration period,
the SSC model might mimic our real universe.  By contrast, if the size
of the LTB region is comparable to the cosmological horizon in the
acceleration period, the behavior of this SSC model might be very different 
from that of our universe. To clarify the situation, 
we investigate which of these two cases corresponds to this SSC model. 
It should again be noted that when the volume of the domain
$\chi<\chi_3$ is dominant in the volume $V_D$ and is contracting, the
volume expansion of the domain $D$ greatly slows down, or the domain
$D$ contracts.  The acceleration of the volume expansion is realized
when the volume of the domain $\chi<\chi_3$ becomes so small that its
contribution to $V_D$ is negligible.  In other words, the acceleration
period appears around the time at which the dust in the domain
$\chi<\chi_3$ collapses to a singularity, and this time is given by
\begin{equation}
t_{\rm ac}:=\frac{8\pi^2}{3}\frac{\rho_0\chi_3^3}{\kappa^{3/2}}.
\end{equation}
At $t=t_{\rm ac}$, the ratio of the areal radius of the boundary 
of a LTB region $a_{\rm e}(t)l$ to the Hubble horizon radius of the EdS
region, 
$H^{-1}(t):=a_{\rm e}/\dot{a}_{\rm e}=3t/2$, 
 becomes
\begin{equation}
\frac{a_{\rm e}(t_{\rm ac})l}{H^{-1}(t_{\rm ac})}
=\left(\frac{2}{3\pi}\right)^{1/3}\kappa^{1/2}
\frac{l}{\chi_3} 
\simeq 0.60\,\kappa^{1/2}\left(\frac{l}{\chi_3}\right).
\end{equation}
Here it should be noted that $\kappa$ has to be almost equal to unity
so that the acceleration of the cosmic volume expansion is realized.
Therefore, the size of the LTB region is comparable to 
the horizon radius of the EdS region in the acceleration  period. 
This means that non-linear general relativistic  
effects are important for the realization of 
the acceleration period. 

\subsection{Two-scale model}

\begin{figure}[H]
  \centering
  \includegraphics[width=0.45\linewidth,clip]{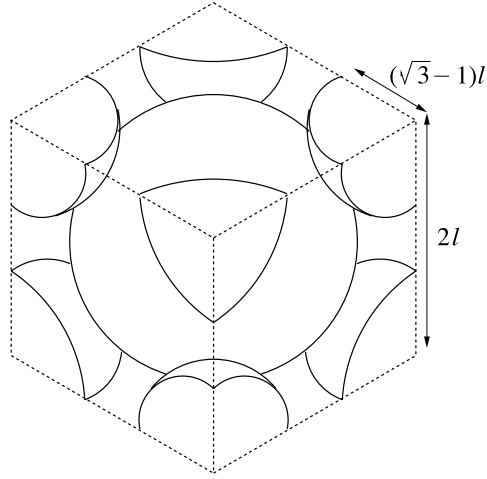}
  \caption{The compact domain $D$ of the two-scale SSC model.}
  \label{fig:two-scale}
\end{figure}

\begin{figure}[H]
  \centering
  \includegraphics[width=0.4\linewidth,clip]{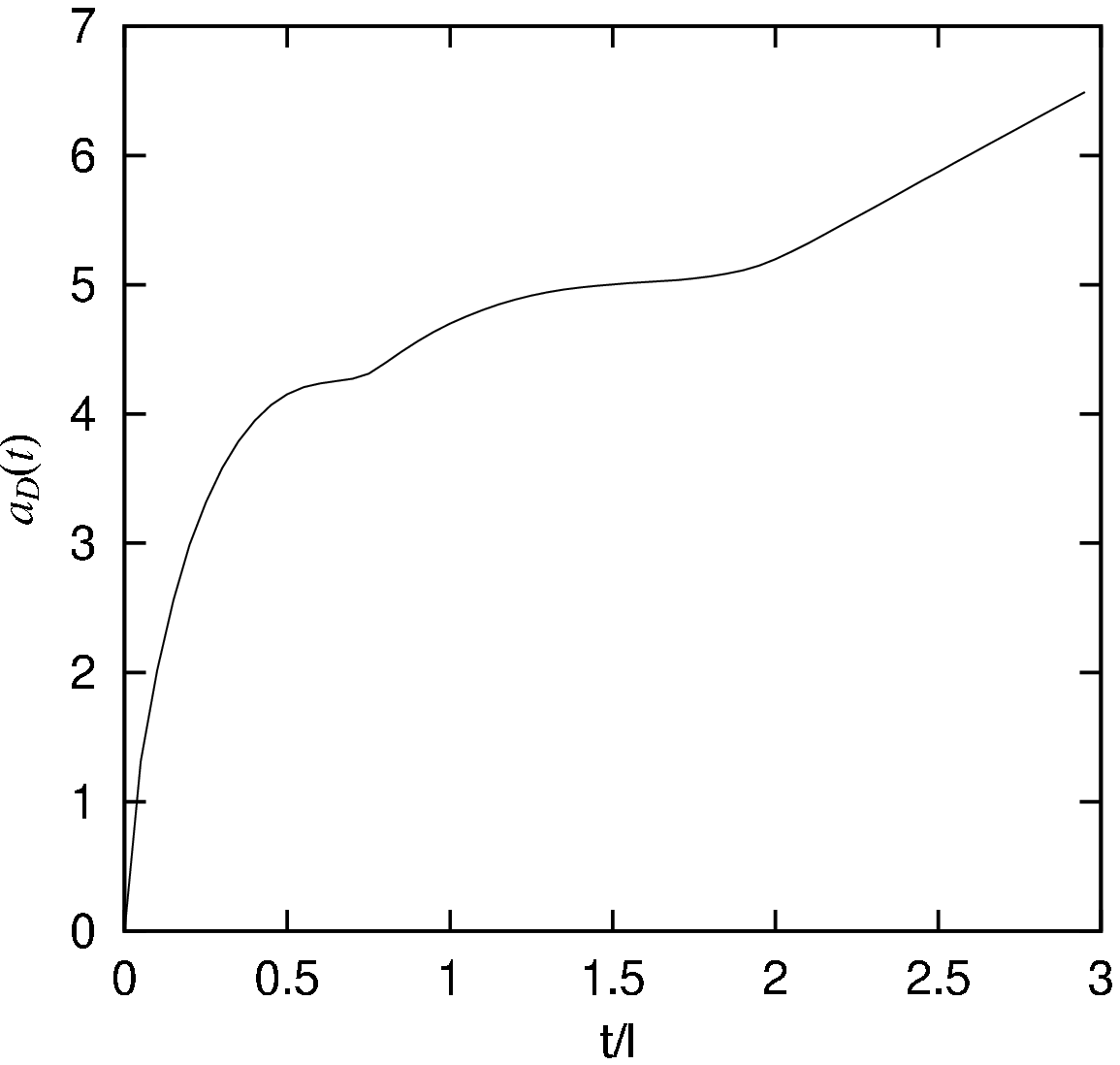}%
  \hspace{0.5cm}
  \includegraphics[width=0.4\linewidth,clip]{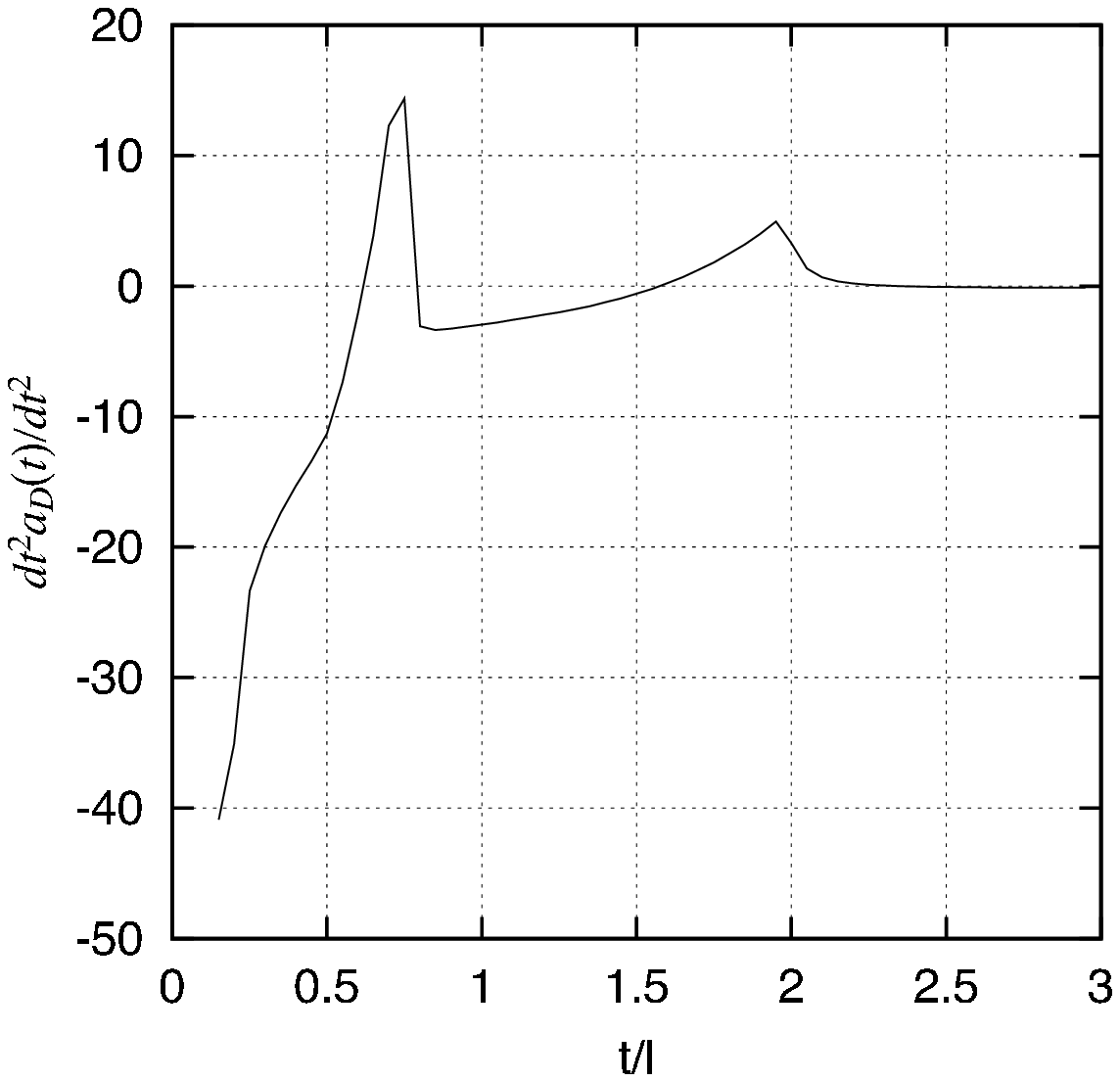}
  \caption{The left panel displays the evolution of the averaged scale
    factor for the two-scale SSC model.  We set $\kappa=0.999688$ in the large
    LTB region and $\kappa=0.999995$ in the small LTB region. Except for
    these values of $\kappa$,
    we used the same parameter values here as in the one-scale SSC model.}
  \label{fig:expansion-two}
\end{figure}

As the second model, we consider the situation that there are two
kinds of LTB regions, one with radius $\chi_{\rm b}=l$ and the
other with radius $\chi_{\rm b}=(\sqrt{3}-1)l$. We assume that the
large LTB region is put in the cubic domain $D$ in the same manner as in
the first model, whereas an eighth of the small LTB region is put on
each vertex of $D$ (see Fig.~\ref{fig:two-scale}). 
In this model, the volume of the compact domain $D$ is given by
\begin{eqnarray}
V_D&=&\left[8
-\frac{4\pi}{3}\left\{1+\left(\sqrt{3}-1\right)^3
\right\}\right]a_{\rm e}(t)l^3 \nonumber \\
&+&4\pi\int_0^{a_{\rm e}(t)l}\frac{Y^2dY}
{\sqrt{1-k(\chi_{\rm L})\chi_{\rm L}^2 }}
+4\pi\int_0^{(\sqrt{3}-1)a_{\rm e}(t)l}\frac{Y^2dY}
{\sqrt{1-k(\chi_{\rm S})\chi_{\rm S}^2 }},
\end{eqnarray}
where $\chi_{\rm L}=\chi_{\rm L}(Y)$ and $\chi_{\rm S}=\chi_{\rm S}(Y)$ 
are the radial coordinates as functions of the areal radius
$Y$ at time $t$ in the large and small LTB regions, through 
Eqs.~(\ref{eq:k>0}) and (\ref{eq:t-solution}), respectively.  We present
one example of the evolution of the scale factor $a_D(t)$ in
Fig.~\ref{fig:expansion-two}.  As expected, there are two acceleration
periods. The first is realized when the small LTB
regions collapse to the singularities, and the second appears 
when the large LTB region collapses. Thus, the
coexistence of various scales of LTB regions can lead to more than one
acceleration period.
\section{Summary and discussion}

Using the stuffed Swiss-cheese model, we investigated how the
cosmic volume expansion is accelerated by inhomogeneities in a
dust-filled universe. We found that the existence of contracting
regions is a necessary
condition to realize the acceleration of the cosmic volume expansion.
Near the time at which the central contracting region collapses to
form a singularity, the acceleration of the averaged scale factor
becomes positive, and the acceleration phase of the spatially averaged
universe appears. Such a situation might be the same as that in the real
universe, 
with structures (e.g. galaxies, clusters of galaxies) 
forming within the contracting regions, 
or there might be large-scale bulk velocity fields due to 
large mass concentrations, in our expanding universe. 
However, acceleration of the cosmic volume expansion is realized only when 
the size of each inhomogeneity is comparable to the cosmological horizon 
scale in the case of the stuffed Swiss-cheese universe model. We see 
from this result that  
this model is very different from the universe 
observed today in the period of accelerating cosmic volume expansion. 
However, for the model with various scales of inhomogeneities, the temporal 
variation of the cosmic volume expansion is nontrivial 
from a theoretical point of view,  
and a complete analysis is left as a future work.
The analysis of the backreaction effect for the
dust-filled universe based on a second-order cosmological
perturbation reveals no signs of acceleration of the volume
expansion\cite{RussH:1997,NambuY:PRD62:2000,KasaiM:0602506:2006}.
Thus, we conclude that the appearance of the acceleration phase in our model is due to
a highly non-linear effect expected to appear beyond third order in the
perturbation analysis.

Here we note that the issue of cosmic acceleration first  
arose from observational data of the distance-redshift relation
and that we have not directly observed the acceleration of the
`volume' expansion. If the universe is homogeneous and isotropic, then
we can conclude from the observational data that the cosmic volume
expansion is indeed accelerating. By contrast, in the inhomogeneous universe model
considered in this paper, the cosmic expansion of the comoving volume
does accelerate, but it is not yet clear whether the distance-redshift
relation is similar to that of the homogeneous and isotropic universe
with accelerating volume expansion.  There are several studies of an
inhomogeneous but isotropic dust-filled universe whose
distance-redshift relation is consistent with the observational
data\cite{CelerierM:AA353:2000,TomitaK:MNRAS326:2001,IguchiH:PTP108:2002,VanderveldRA:0602476:2006}.
In these models, there is no contracting region, and therefore the
cosmic volume expansion does not accelerate. 
Conversely there is a possibility that the acceleration of the cosmic
volume expansion due to inhomogeneities does not lead a
distance-redshift relation similar to that obtained in the observation
of Type Ia supernavae. Investigating  this point is also a future work.

\section*{Acknowledgements}

We are grateful to our colleagues in the astrophysics and gravity group of
Osaka City University for helpful discussions and criticism.

\end{document}